\documentstyle[11pt,aaspp4]{article}

\newcommand{\etal}{{\frenchspacing\it et al. }}

\newcommand{\lsim}{\hbox{ \rlap{\raise 0.425ex\hbox{$<$}}\lower 0.65ex\hbox{$\sim$} }}
\newcommand{\gsim}{\hbox{ \rlap{\raise 0.425ex\hbox{$>$}}\lower 0.65ex\hbox{$\sim$} }}

\begin{document}

\title{The Northern Sky Optical Cluster Survey I:, \\
  Detection of Galaxy Clusters in DPOSS}

\author{R.R. Gal\altaffilmark{1}, R.R. de Carvalho\altaffilmark{2}, S.C. Odewahn, S.G. Djorgovski}
\affil{Palomar Observatory, Caltech, MC105-24, Pasadena, CA 91125}
\and

\author{V. E. Margoniner}
\affil{Observat\'orio Nacional, Rua Gal.  Jos\'e Cristino,
77 -- 20921-400, Rio de Janeiro, RJ, Brazil}

\altaffiltext{1}{Email: rrg@astro.caltech.edu}
\altaffiltext{2}{Observat\'orio Nacional, Rua Gal.  Jos\'e Cristino,
77 -- 20921-400, Rio de Janeiro, RJ, Brazil}

\begin{abstract}
The Northern Sky Optical Cluster Survey is a project to create an objective catalog of galaxy clusters over the entire high--galactic--latitude Northern sky, with well understood selection criteria. We use the object catalogs generated from the Digitized Second Palomar Sky Survey (DPOSS, Djorgovski \etal 1999) as the basis for this survey. We apply a color criterion to select against field galaxies, and use a simple adaptive kernel technique to create galaxy density maps, combined with the bootstrap technique to make significance maps, from which density peaks are selected. This survey attempts to eliminate some of the subjective criteria and assumptions of past surveys, including detection by eye (Abell 1958, ACO 1989) and assumed luminosity functions and cluster profiles (PDCS, Postman \etal 1995). We also utilize more information (especially colors) than the most similar recent survey, the APM (Dalton \etal 1992). This paper presents the details of our cluster detection technique, as well as some initial results for two small areas totaling $\sim60$ square degrees. We find a mean surface density of $\sim 1.5$ clusters per square degree, consistent with the detection of richness class 0 and higher clusters to $z\sim0.3$. In addition, we demonstrate an effective photometric redshift estimator for our clusters.

\end{abstract}

\clearpage
\section{Introduction}

Clusters of galaxies are the largest bound systems in the Universe
and, as such, have been used to study how matter is distributed over extremely 
large scales.  They provide useful constraints for theories of 
large-scale structure formation and evolution, and are valuable (possibly 
coeval) samples  for studying galaxy evolution in dense environments.  Studies
of the cluster two-point correlation function and the power spectrum are 
important probes of large scale structure and the scenarios of its formation.  
Correlations between optically and X--ray selected (e.g., ROSAT, Chandra, etc.) clusters are also of considerable scientific interest, and will help us better understand the various selection effects present in both types of cluster samples.

Most of the optical studies to date have been limited by the statistical 
quality of the available cluster samples. For instance, the Abell catalog
(Abell 1958, ACO 1989) suffers from a non-objective selection process, poorer plate material, a bias towards centrally concentrated clusters, especially those with cD galaxies, a relatively low redshift cutoff  ($z \sim0.15$; Bahcall \& Soneira 1983), and strong plate--to--plate sensitivity variations. Still, many far--reaching  cosmological conclusions have been drawn from it (i.e. Bahcall \& West 1992), although later studies have sometimes shown these to be flawed. Additionally, projection effects are a serious and difficult to quantify issue. These effects resulted in early findings of excess large--scale power in the angular correlation function (Bahcall \& Soneira 1983), and later attempts to disentangle these issues relied on models to decontaminate the catalog (Sutherland 1988, Olivier \etal 1990). Unfortunately, some of these will plague any optically selected cluster sample, including our own, but objective selection criteria and a strong statistical understanding of the catalog can mitigate their effects.

Other catalogs preceding ours have been generated using objective means (APM, Dalton \etal 1992; EDCC, Lumsden \etal 1992). The APM group, for instance, used 
digitized $J$ (blue) plates from the Southern Sky Survey; the use of a single, blue band provides no color information to distinguish galaxy types, and is a poor choice for cluster detection because clusters are better delineated by redder, early-type galaxies in the redshift range we probe ($z<0.4$). The survey presented here utilizes at least one color (two filters), and is based on high resolution, low scattered light scans of photographic survey material, providing excellent classification and photometry. Additionally, this survey will eventually cover the entire high--galactic--latitude northern sky, a much larger area than some prior efforts (such as the EDCC).

This paper presents the initial results of an optical cluster survey based on DPOSS data, from two fields covering $\sim60$ square degrees.
We use the 3 bands ($JFN$) of the DPOSS survey (Djorgovski \etal 1999) calibrated with extensive CCD data into the Gunn $gri$ system. From this data, we construct color-color diagrams for galaxies detected in all 3 bands. Predicated on the morphology-density relation (Dressler 1980), where redder, early--tpye galaxies (E/S0) are preferentially found in higher density regions, we select galaxies that are more likely to appear in clusters based on their colors. Alternatively, this can be viewed as a means of excluding galaxies that are more common in the field. After the color cuts are applied, we use the adaptive kernel technique (Silverman 1986) to produce a surface density map of galaxies. A bootstrap method is then used to construct the corresponding significance map, from which density enhancements are selected using the FOCAS peak finding algorithm. While previous surveys have tested, and sometimes utilized, standard photometry packages for cluster detection (Shectman 1985, Lumsden \etal 1992), the novelty of our technique lies in the use of a color--selection technique, over a very large area of the sky, with an adaptive kernel to generate the input galaxy density maps.

In the next section of this paper, we briefly describe the input data derived from DPOSS. Calibration and classification issues are briefly discussed, followed by our color selection procedure.  The third section covers the density mapping and peak selection. The fourth section presents our initial results for two DPOSS fields ($\sim60$ square degrees), and compares our results with those from previous surveys. In the fifth section, we discuss measurements of cluster properties directly from the calibrated plate photometry, including our attempts to estimate cluster redshifts using colors. In the final section we discuss future work and present our conclusions.

\section{Input Data and Catalog Preparation}

\subsection{Input DPOSS Data}

The Northern Sky Optical Cluster Survey is based on data from the Digitized Second Palomar Sky Survey (DPOSS, Djorgovski \etal 1999), a digitization of the POSS--II  three--band photographic survey of the entire Northern sky (Reid \etal 1991). Each plate covers $6.6^{\circ} \times 6.6^{\circ}$ on the sky, with neighboring plates overlapping each other by $\sim1.6^{\circ}$. The plates are scanned at STScI (lasker \etal 1996), with $1''$ pixels. The typical seeing in the digitized data (combining both the effects of telescope seeing, and the scanning process) is $\sim 2''$. The digitized data are processed into catalogs at Caltech using the SKICAT system (Weir \etal 1995c). The end result of plate processing is a catalog of all objects detected down to the limiting magnitude of the plate ($g_J\sim21.5,r_F\sim20.5,i_N\sim20.0$), with various photometric, positional and shape parameters. This data is collected from all processed plates into the Palomar Norris Sky Catalog (PNSC), expected to contain $\sim 50x10^6$ galaxies and $>2x10^6$ stars. Classification of each object is performed by a decision tree, which has been shown to be $> 90\%$ accurate for objects brighter than $r=20.0^m$ (Weir \etal 1995a; Odewahn \etal 1999). The data are calibrated into the Gunn system using an extensive collection of CCD imaging data obtained at the Palomar $60''$ telescope. Typical photometric errors for extended sources at in each band are $\sigma_g\sim.21^m,\sigma_r\sim.18^m,\sigma_i\sim.37^m$ at 19th magnitude (Odewahn \etal 1999).

Each field in each band is processed individually. The three resulting catalogs are cross--matched to create a composite list of objects for the field. We require a detection in both the $J$ and $F$ bands so that we can measure at least one color. $N$ detections are not required because the $N$ data are not as deep as the other two bands, and suffer from large plate sensitivity variations. In practice, however, the relatively bright $r$ magnitude limit imposed to maintain accurate classification results in most objects ($\sim98\%$) being detected in all three bands.

Finally, those areas on the plate containing saturated objects are masked. These areas often contain large numbers of falsely identified galaxies, as the plate processing software handles large, bright objects improperly. The masked areas, on average, cover $7\%$ of the plate. Figure 1 shows Field 447, with the masked areas enclosed in boxes. These areas are not used in cluster detection.

In this paper, we present results for two fields: 447 ($14^{h} 30^{m} +30^{\circ}$) and 475 ($01^{h} +25^{\circ}$). These fields were chosen because they are at relatively high galactic latitude ($+67^{\circ}$ and $+40^{\circ}$) where the effects of dust are expected to be small (but see the discussion in section 5), and because scans in all three bands were available when this project was started.

\subsection{Selecting Cluster Galaxies Using Colors}
The morphology-density relation (Dressler 1980) has been observed in galaxy clusters at low redshift for nearly two decades. Late--type galaxies are dominant in the field population, whereas early--type galaxies are preferentially seen in high density regions. Therefore, any technique that can eliminate field (i.e. late--type) galaxies on the basis of some simple observable parameter will enhance the contrast of galaxy clusters relative to the background.

Dressler \& Gunn (1992) presented photometric data for seven rich clusters. They show that cluster galaxies follow a well--defined sequence in the $(g-r)$ vs. $(r-i)$ color-color space, whereas field galaxies occupy a much larger area in this space. By defining a locus of ``cluster galaxies'', and using only those objects which satisfy this color criterion, it is possible to preferentially discard field galaxies from the sample. Such a procedure was utilized effectively by Odewahn \& Aldering (1995) in their cluster detection program. The Palomar Distant Cluster Survey (Postman \etal 1995), using deep, multi--color CCD data, also employed colors to detect clusters and estimate their redshifts. However, No other large (all--sky), systematic cluster survey has been able to use color information for this purpose, as the data simply did not exist.

Figure 2 shows the $(g-r)$ vs. $(r-i)$ diagram for galaxies which are
matched in the three plate catalogs ($JFN$) for Field 475.  We use only objects classified as galaxies on the $J$ or $F$ plate (whichever has better seeing), with calibrated magnitudes  $r < 20.0^m$, where our classification accuracy is $> 90\%$ (Weir \etal 1995b; Odewahn \etal 1999). In practice, this imposes an approximate redshift limit of $z\sim0.3$ on our cluster detection.

Initially, we defined a strict locus of cluster galaxies in the $(g-r)$ vs. $(r-i)$ space, similar to the prescription given by Dressler \& Gunn (1992). However, inspection of our color--color diagrams shows that the $i$ photometry is poor, resulting in large scatter in the $r-i$ color. We have therefore chosen to make color cuts in the $g-r$ space only. The chosen cuts in $g-r$ have two motivations. On the blue side, $(g-r)<0.3$, we are excluding blue field galaxies at low $z$. On the red side, $(g-r)>1.3$, we are excluding objects that have a high likelihood of being either misclassified stars, or at high redshift. These color limits are marked in Figure 2. Clearly, a significant fraction ($\sim30\%$) of galaxies in our catalog lie outside our chosen $g-r$ range.  We have also tested different color selection criteria, and find that only the lowest significance clusters, usually near the plate edges, are affected. In addition, the $k$--corrected colors for both E and Scd
galaxies are shown, for the redshift range $0.0<z<0.8$, for the Gunn filters used to calibrate the data. The differences between the actual plate filter/emulsion combination and the CCD filters used for calibration do not introduce any significant effects. The sharp turn in the $k$--correction curves occur at $z=0.4$, where the 4000{\AA}  break passes from the $g$ to the $r$ band. Not surprisingly, we can see evidence for the locus of cluster galaxies along the $k$--correction track for early-type galaxies, albeit with large scatter.
From this figure, it is clear that we are seeing galaxies out to $z=0.4$, although we are very incomplete above $z=0.25-0.3$. This is not surprising, as an $L_{\star}$ early-type galaxy will have $r \sim20^m$ at $z\sim0.3$ for any reasonable cosmology. 

For comparison, Figure 3 shows the results of a similar analysis of a number of Abell clusters imaged at the Palomar $60''$ telescope. The cluster galaxy sequence is very well delineated, but is clearly truncated for $z>0.2$, as is well known (Abell \etal 1989).

\section{Galaxy Density Maps and Cluster Detection}

\subsection{Density and Significance Maps}
Once a catalog of galaxies meeting our classification, magnitude, and color criteria has been created, we proceed to detecting candidate clusters. We have chosen to use the Adaptive Kernel (AK) technique (Silverman 1986). This technique uses a two--stage process to produce a density map. First, at each point $t$, it produces a pilot estimate $f(t)$ of the galaxy density at each point in the map. We use a fixed Epanechnikov kernel, 
\begin{equation}
{K(t)=\cases{{3\over4\,\sqrt5}\left(1-{1\over5}t^2\right) & $-\sqrt5 \le t \le \sqrt5$; \cr
0,&otherwise. \cr}}
\end{equation}
 of 15' width. Based on this pilot estimate, it then applies a smoothing kernel whose size changes as a function of the local density, with a smaller kernel at higher density. This is achieved by defining a {\it local bandwidth factor}:
\begin{equation}
{\lambda_i = [f(t)/g]^{-\alpha},}
\end{equation}
where $g$ is the geometric mean of $f(t)$. We use a sensitivity parameter $\alpha = 0.5$, which results in a minimally biased final density estimate, and is simultaneously more sensitive to local density variations than a fixed--width kernel (silverman 1986). This is then used to construct the adaptive kernel estimate:
\begin{equation}
{\hat f(t) = n^{-1}\sum_{i=1}^nh^{-2}\lambda_i^{-2}K\{h^{-2}\lambda_i^{-2}(t-X_i)\}}
\end{equation}
where $h$ is the bandwidth. The choice of bandwidth is discussed below.

This technique has many advantages over methods used by other surveys. First, the two--step process significantly smooths the low density regions, without affecting the high density peaks. The smaller kernel size at high density means that multiple clusters in high density regions (e.g., filaments) are still separated. This technique may be more robust than percolation or friends--of--friends algorithms, which may link clusters in regions of high galaxy density, since they employ a fixed linking length.  Second, the AK does not require an assumed luminosity function or radial profile for the cluster, as did the matched filter of the PDCS. With our plate data, the range in magnitude covered is small ($16\le m_r\le 20$), so that fitting a luminosity function is a precarious endeavor. More importantly, perhaps, visual inspection of our candidate clusters shows that many (perhaps most) do not appear symmetric, so that any radially symmetric profile, assumed by some techniques like the matched filter (Postman \etal 1996) could produce biased results. We are attempting to avoid any bias towards rich, relaxed, evolved clusters (which may occur if clusters are selected from  X--ray surveys). This results in a catalog in which some clusters will have different physical properties than those discovered by other techniques and at other wavelengths. 

Our density maps are created using $1'$ pixels, resulting in a final map size of $360 \times 360$ pixels for each field. The kernel itself has a bandwidth comparable to the Abell radius of a cluster at $z\sim.15$, which is expected to be the median redshift of our cluster candidates. This combination of pixel size and bandwidth prevents us from over--resolving clusters into structural components, while maintaining a reasonable number of galaxies per pixel. Once a density map has been constructed, we use the {\it bootstrap} technique (Press \etal 1992) to build a significance map. This technique uses the input data set,  $C_{(0)}^{S}$, containing {\it N} points, to generate synthetic data sets $C_{(1)}^{S}$, $C_{(2)}^{S}$, ... with the same number of points {\it N} as the input set. These {\it N} points are selected at random from the input set, with replacement. This means that the same point can be selected more than once for each synthetic data set. We generate 500 synthetic data sets, each resulting in a density map $D_{i}^{S}(j,k)$, with map number $i=1, 500$ and pixels $j,k=1, 360$. We define a mean map, $\cal M$, as:

\begin{equation}
{\cal M}(j,k)={1 \over 500}{\sum_{i=1}^{500}{D_{i}^{S}(j,k)}}
\end{equation}

\noindent {and an analogous map ${\cal M}2$:}

\begin{equation}
{\cal M}2(j,k)={1 \over 500}{\sum_{i=1}^{500}{[D_{i}^{S}(j,k)]^2}}
\end{equation}

\noindent {From these, we form a final significance map ${\cal S}$:}

\begin{equation}
{\cal S}(j,k) = {\cal M}(j,k) - 3({\cal M}^{2}(j,k)-{\cal M}2(j,k))^{1 \over 2} 
\end{equation}

Density maps and corresponding significance maps for fields 447 and 475 are shown in Figures 4 and 5. One can see that the structures seen in the significance map are also visible in the original density map.

\subsection {Selecting Cluster Candidates}

The significance map constructed using the bootstrap technique is then used to identify peaks in the galaxy distribution, which we mark as candidate galaxy clusters. We utilize the FOCAS peak-finding algorithm (Jarvis \& Tyson 1981) to detect enhancements with $ \ge 3.0 \sigma {_F}$ significance, where $\sigma {_F}$ is the $rms$ of the background in the significance map. In practice, this is a very liberal detection threshold, in the sense that we accept a larger number of false detections in return for both enhanced completeness at higher redshift and lower richness. We also impose a minimum area requirement of 66 pixels, which corresponds to a circular region of $\sim0.8$ Mpc observed at $z = 0.15$. This yields 47 and 38 candidate clusters in fields 447 and 475, respectively. To quantify our false detection rate (which may be large at low significance), we have undertaken a spectroscopic follow--up campaign to measure redshifts for all the cluster candidates in these two fields, with results to be presented in a future paper.

The candidate clusters are marked in Figures 4 and 5, with Abell clusters shown in green, and new clusters in white. We successfully recover all known Abell clusters in these fields, and find a vastly larger number of new candidates. Clearly, Abell identified, in general, only the most significant density enhancements, although some rich clusters (such as candidate 31 in Field 475, shown in Figure 6) were still missed.

\section {Estimating Cluster Properties}

After detection from the significance map, some simple properties (such as size) of the clusters are measured from the original density map. Properties of the galaxy population of each cluster are derived from the input DPOSS galaxy catalogs. This helps to eliminate spurious detections, and allows us to compare our catalog with pre-existing catalogs. In this paper, we address properties measured directly from the plate data, while later papers will utilize additional CCD imaging and spectroscopic data obtained for the cluster candidates in these two fields for more detailed studies.

\subsection {Photometric Redshift Estimation}

From the plate data, we wish to measure the redshift and richness of each cluster candidate. First we estimate the redshift assuming that each cluster candidate contains a single cluster, at one redshift, and is dominated by early-type galaxies. We count the number of galaxies as a function of color, $N_{g-r}$, and the number as a function of $r$ magnitude, $N_r$, inside the cluster area, as determined by FOCAS from our galaxy density maps. The background galaxy color and magnitude distributions, $N_{bg,g-r}$ and $N_{bg,r}$ are determined from a large ($\sim146$ square degrees) galaxy catalog around the North Galactic Pole, constructed from DPOSS data (Odewahn \etal 1999). This distribution, scaled to the appropriate area, is then subtracted from the color and magnitude distributions of each candidate cluster, and the median $g-r$ color and mean $r$ magnitude of the remaining galaxies calculated. 
Figure 7 shows the $N_{g-r}$ distribution for a typical candidate before  subtraction, the scaled $N_{bg,g-r}$ background distribution, and the final $N_{corr,g-r}$ distribution after background correction. As expected, the background galaxy counts are clearly bluer than the cluster galaxies. We also experimented with the use of locally derived background estimates. We found that the closer to the cluster the local background was determined, the lower the estimated cluster richness. This effect continues out to very large radii, and prompted the use of a global background estimate, which has been questioned in the past (Lumsden \etal 1992).

This procedure was performed for 46 known Abell clusters with measured redshifts $z_{spect} > 0.05$, from 10 plates. However, in this case, we set the radius to the Abell radius ($R_{g-r} = 1.7^{\prime}/z_{g-r}$) (Abell 1958) to measure the colors and magnitudes. We find that both the colors and magnitudes are equally well correlated with measured spectroscopic redshifts. We derive an empirical relation between redshift, median $g-r$ color, and mean $r$ magnitude using a bivariate least--squares fit:
\begin{equation}
{z_{phot} = 0.1074 \times r_{mean}  + 0.1317 \times (g-r)_{med} - 1.9706}
\end{equation}

The $rms$ of $z_{spect} - z_{phot}$ is $\small {\Delta}z=0.0165$. Figure 8 shows the photometrically estimated redshift against the spectroscopically measured redshift for all 46 clusters. Open circles show those clusters with only one galaxy with a spectroscopic redshift, and filled circles are clusters with two or more spectroscopic redshifts. 

In practice, this must be an iterative procedure, because we do not initially know the redshift, and therefore the Abell radius, for our cluster candidates. Once a redshift is estimated, we then determine the Abell radius ($R_{phot}
= 1.7^{\prime}/z_{phot}$) for each candidate, and recenter on the median location of galaxies inside this radius. We then repeat the background correction and color and magnitude measurements inside $R_{phot}$, centered on the new location, and estimate the redshift anew. These steps are repeated until the redshift converges; $\small {\Delta} z_{phot} < 0.01$.

The complete iterative procedure was also performed on the 46 clusters with measured redshifts. As expected, starting the process with the incorrect Abell radius increases our errors. The final $rms$ of $z_{spect} - z_{phot}$ is $\small{\Delta}z=0.026$, with a mean offset of only $-0.0016$. This error includes all random and systematic errors, such as  plate--to--plate photometry, classification, background correction, and even incorrectly measured spectroscopic redshifts. Approximately 50\% of the 46 clusters used have only one galaxy with a measured redshift. This result does not vary strongly with the radius used ($R_{Abell}$ or $0.5\times R_{Abell}$). In the future, as we gather more data, we will recalculate our photometric redshift estimator. Additionally, we have ignored clusters with $z_{spect} < 0.05$, as these are too extended for our technique to handle properly, and our photometric calibration for bright galaxies is currently too poor.

We find the small errors from this photometric redshift technique to be remarkable, particularly in light of the data we are using. The data is taken from 10 plates in various parts of the sky, and therefore encompasses all of our photometric and classification errors, as well as varying amounts of extinction. The photometry is taken from photographic plates, where the quality is much lower than for CCDs. Additionally, the galaxy populations of these clusters must also vary; the fraction of bluer, late--type galaxies is not constant with richness or redshift. 
In comparison, the APM group, utilizing a richness-dependent, magnitude--based redshift estimator, achieved similar errors over only a much smaller redshift range ($0.04<z<0.1$, Dalton \etal 1994).

\subsection {Richness}
We have attempted to estimate the richness of our cluster candidates in numerous ways. We tested Abell's prescription (counting galaxies between $m_3$ and $m_3+2$ mag, where $m_3$ is the magnitude of the third brightest galaxy), with and without a background correction applied, as well as simply counting all galaxies in the range $16.0^m < r < 20.0^m$. The richnesses of clusters with spectroscopic redshifts was measured from our catalog data, and compared directly to the data from Abell (1958). Rather than use richness class (which has a small range, is not continuous, and is dominated by poor clusters), we used the number of galaxies in the cluster. We find that there is a good correlation between our richness measures and those from Abell for clusters at $z < 0.1$, but at higher $z$, we consistently underestimate the richness. This effect is certainly not new; Scott (1953) demonstrated the redshift dependence of such a simple--minded richness measure. AS expected, the degree of underestimation worsens as $z$ increases. This is due to the $m_r = 20.0$ cut imposed due to classification. By $z=0.1$, we are no longer able to sample a reasonable part of the luminosity function of the cluster, and at $z=0.25$, only the very brightest cluster members are visible. 

We postpone discussion of richness estimation for a later paper. We plan to carry out further tests using objects fainter than $r=20.0^m$, perhaps ignoring classification, as was done by Croft \etal (1997). At fainter magnitudes, the objects in a putative cluster should be predominantly galaxies, so that including them in a richness estimate may be feasible. This would force us to increase the limiting richness for a complete sample at higher redshift. We may also assume a Schechter luminosity function, fitting only the bright end, and estimate a cluster's richness from the normalization of the LF. It should be noted that the APM survey did not encounter this problem because they limited themselves to $z<0.1$ for estimating richness and redshift.

In Tables 1 and 2 we provide the relevant information about the selected 
cluster candidates for fields 447 and 475. Column (1) gives the identification number of the candidate; columns (2) and (3) show the right ascension and 
declination coordinates for J2000.0; column (4) shows the 
redshift estimated from our photometric technique; and column (5) identifies which are Abell clusters. Where no estimated redshift is given our estimator failed to converge.

\subsection{Extinction}
An important consideration in the construction of an all--sky cluster catalog is extinction from dust in our Galaxy.
Because the two fields considered in this paper are at high galactic latitude, the extinction correction is negligible. The mean extinction E$(g-r)$ derived from the Schlegel \etal (1998) maps is only $0.01^{m}$ and $0.05^{m}$ for fields 447 and 475, respectively. Nevertheless, our final all-sky catalog will use extinction corrected magnitudes and colors to derive cluster properties. It is also instructive to compare the $100 \mu$ IRAS map, and the galaxy surface density map, for another DPOSS field, number 479, shown in Figure 9. It is clear that galaxies are extincted out of our detection threshold, creating false large-scale structure. This cannot be avoided; it is impossible to correct the color of a galaxy that is not detected. This effect has been largely ignored by past cluster surveys, which only corrected magnitudes of detected galaxies , or were restricted to areas of low extinction. (Maddox, Efstathiou \& Sutherland 1996). This can have dire consequences for measurements of large scale structure, introducing false power on all scales. Clusters are detected where the galactic extinction is low, and lost where the extinction is high. This non--uniform detection threshold across the sky must be accounted for when measuring the cluster--cluster correlation function, for instance.

\section{Discussion and Conclusion}

We have presented a simple but robust technique for generating an objective catalogue of galaxy clusters to $z\sim0.3$ over the entire high--galactic--latitude Northern sky, using DPOSS data. With the multicolor nature of our data, and our demonstrated ability to measure redshifts photometrically, many scientific problems can be addressed. Future papers will present cluster catalogs over large areas of the sky, stitched together seamlessly from individual field catalogs (thus avoiding plate edge effects). Fundamental astrophysical questions will be addressed, including the cluster--cluster correlation function and the evolution of galaxy populations in clusters. Our observed cluster density, $\sim1$ per square degree, is consistent with the detection of richness class 0 and higher clusters to $z\sim0.3$. It is approximately 50\% higher than the space density of clusters found by APM (Dalton \etal 1992).

Additionally, we have undertaken an extensive follow-up campaign, including deep ($r_{lim} = 23.0^m$) 3-band ($gri$) CCD imaging at the Palomar $60''$ telescope, and multislit spectroscopy at the Palomar $200''$ Hale telescope. Every candidate in our two test fields will have both follow-up imaging and spectroscopy, with results to be presented in later papers. While this data will not free our sample of selection effects and biases (most notably due to overlapping clusters), it  will allow us to provide an unprecedented statistical understanding of our cluster catalog. The spectroscopy will provide redshifts of $\sim80$ candidate clusters in the two fields described in this paper. We will be able to directly measure our false detection rate and the redshift distribution of our clusters. The CCD imaging will provide detailed constraints on our photometric and classification errors, in addition to enabling studies of cluster luminosity functions, galaxy populations, etc. The data may also be used in searches for clusters at higher redshift.

Finally, we note that the next large--area galaxy cluster catalogs will not be produced for some years. These will likely be redshift--selected samples from the 2dF and SDSS redshift surveys. The next complete photometrically selected sample, from the SDSS imaging survey, is in the somewhat more distant future. Until then, we hope that this modern catalog of galaxy clusters will be useful in broad--ranging, multi--wavelength studies by ourselves and other investigators. Upcoming and recent space missions, such as CXF and XMM, will also benefit greatly from this new resource.

\acknowledgments

RRG was supported in part by an NSF Fellowship, NASA GSRP NGT5-50215, and a Kingsley Fellowship. The Norris Foundation has provided generous support for the DPOSS project and creation of the PNSC. Eastman Kodak has provided the plates and additional support for POSS-II. We owe a debt of gratitude to the entire POSSII survey team for their dedication to the long and arduous process of taking over 2700 photographic plates. We are also indebted to the plate scanning team at STScI, especially Barry Lasker, for their efforts.  RRG thanks R. Brunner for help with photometric redshifts. RRdC would like to thank S. Schechtman for very useful suggestions during this project. M. Pahre has also participated in enlightening conversations, and H. Capelato kindly provided his adaptive kernel software. Finally, we thank the referee for the rapid response and helpful comments.

\clearpage

\begin{figure}
\plotone{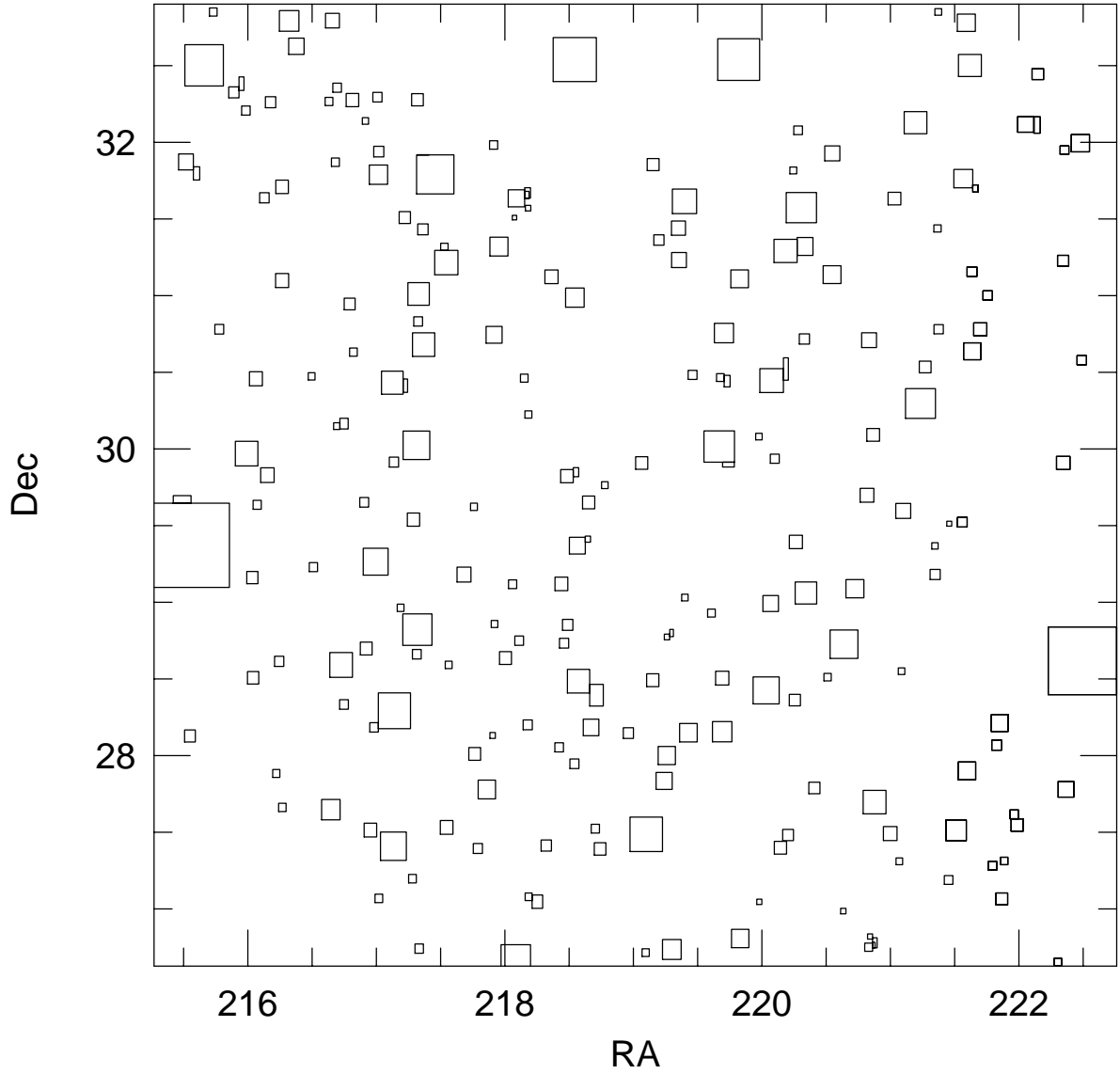}
\caption{The areas masked in DPOSS Field 447 due to saturated objects. These areas are not used in cluster detection.}
\end{figure}

\begin{figure}
\plotone{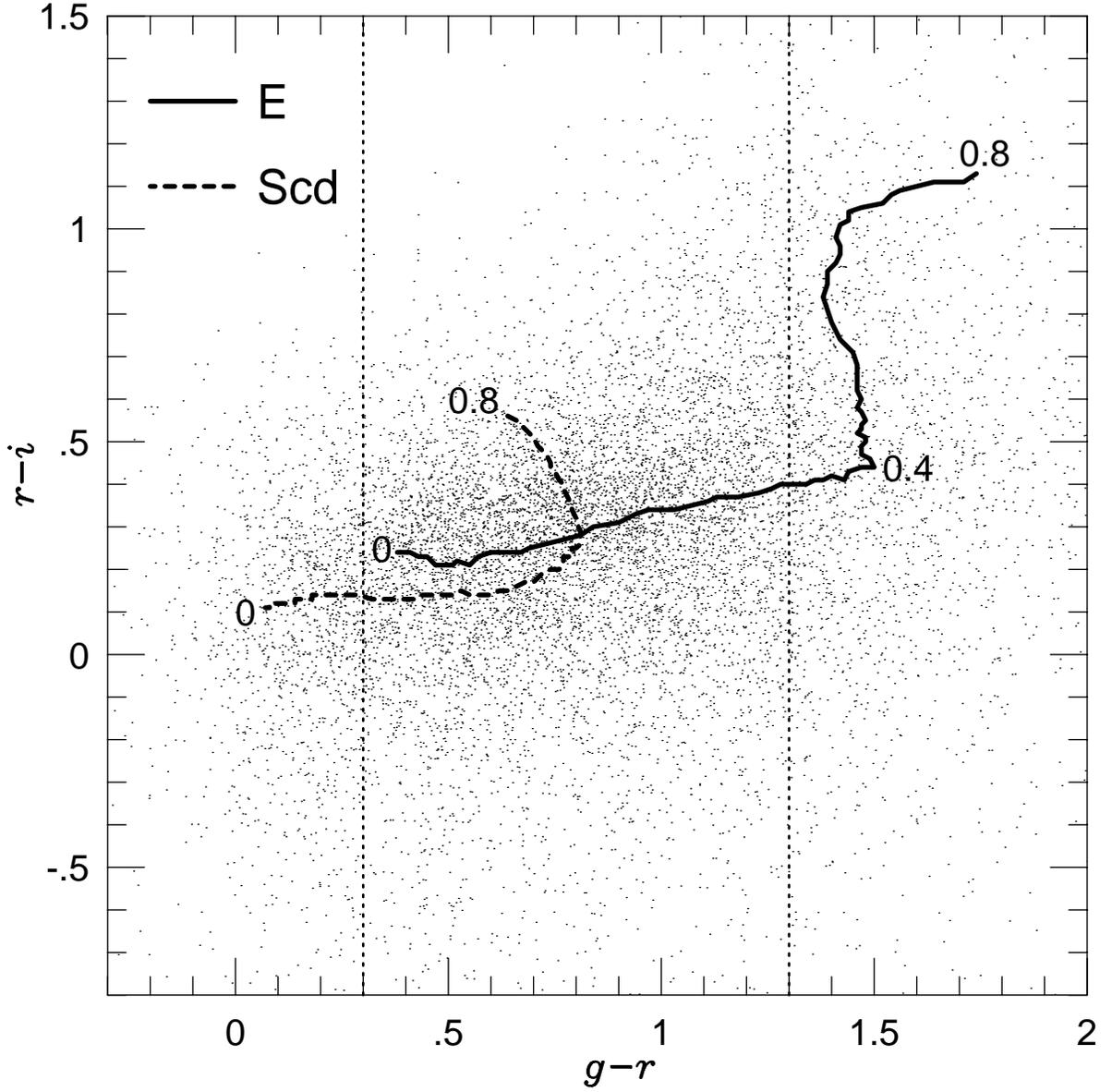}
\caption{The $(g-r)$ vs. $(r-i)$ diagram for galaxies in Field 475. The $k$--correction curve for E galaxies is shown as the solid line, and for Scd galaxies as the dashed line, with some redshifts marked. The vertical dotted lines denote our color cuts, which enclose $\sim70\%$ of the objects.}
\end{figure}

\begin{figure}
\plotone{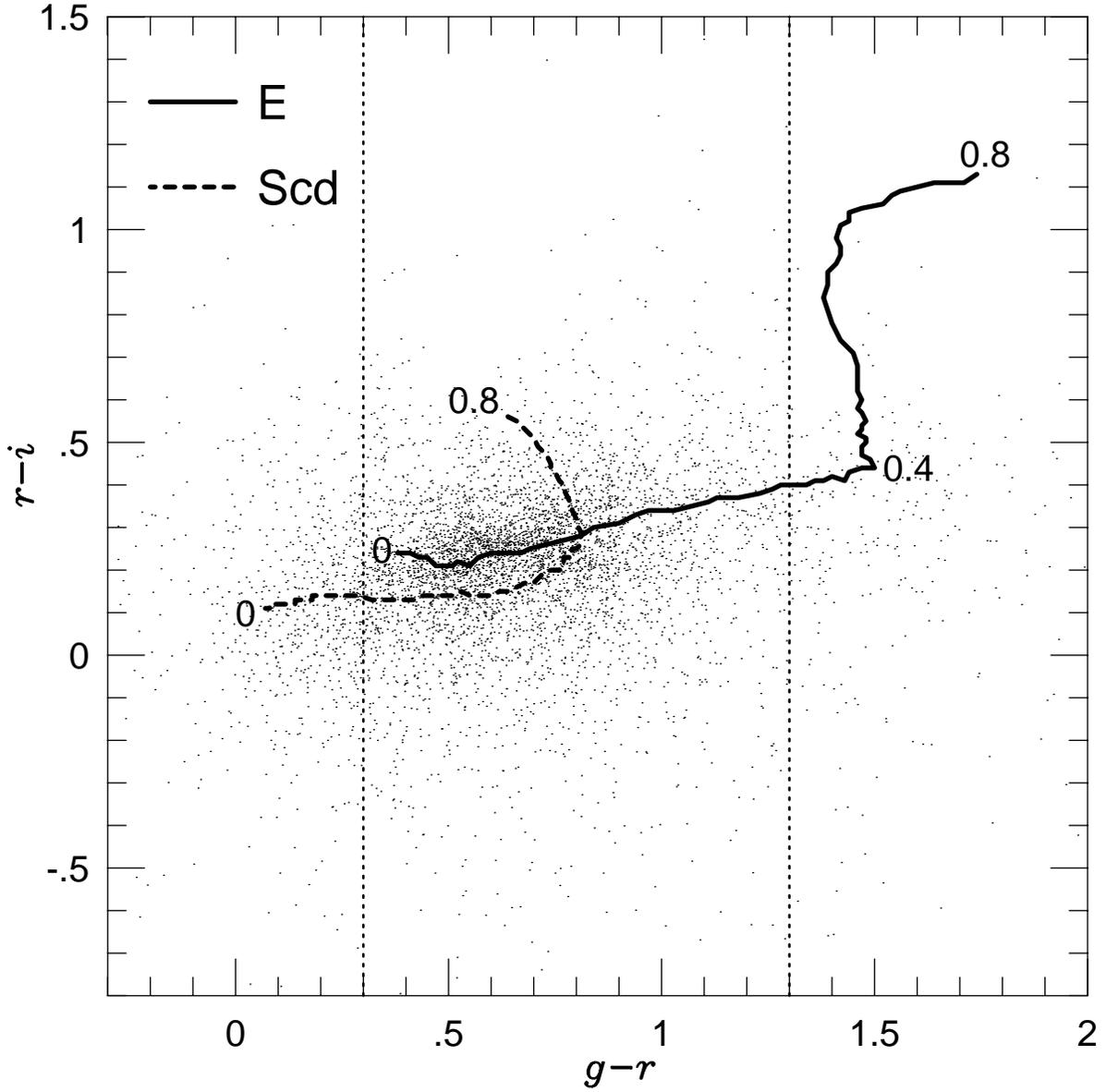}
\caption{The $(g-r)$ vs. $(r-i)$ diagram for galaxies in Abell clusters observed at the Palomar $60''$ telescope. The $k$--correction tracks from Figure 2 are shown for comparison.}
\end{figure}

\begin{figure}
\plotone{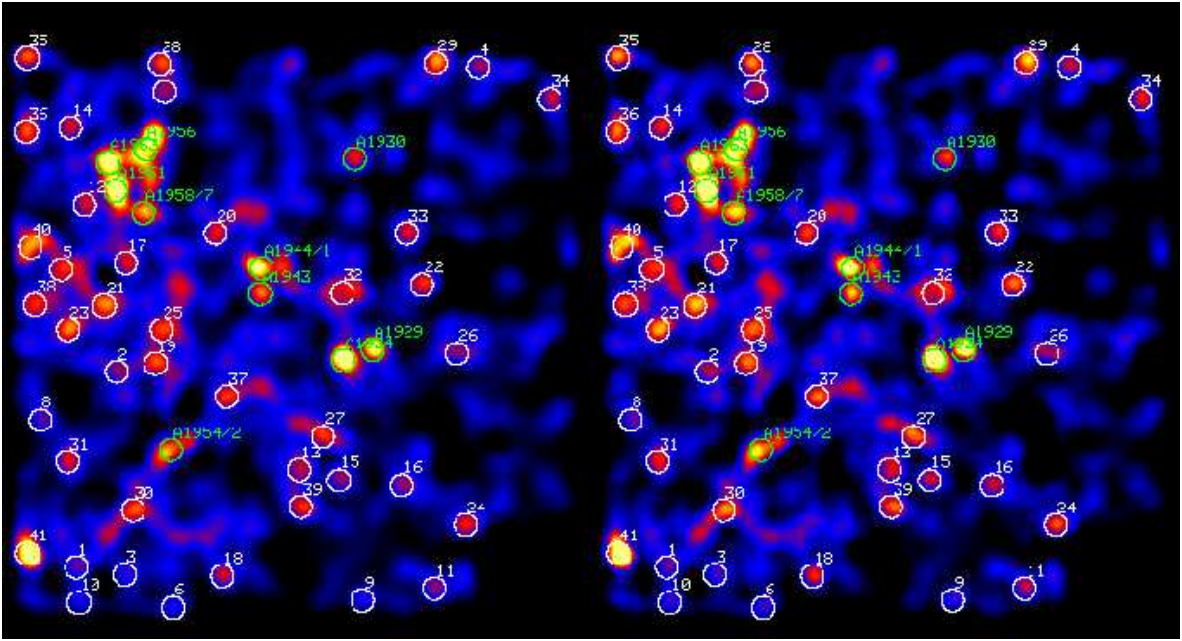}
\caption{Density and significance maps for DPOSS Field 447. Abell clusters are marked in green, and new candidate clusters in white.}
\end{figure}

\begin{figure}
\plotone{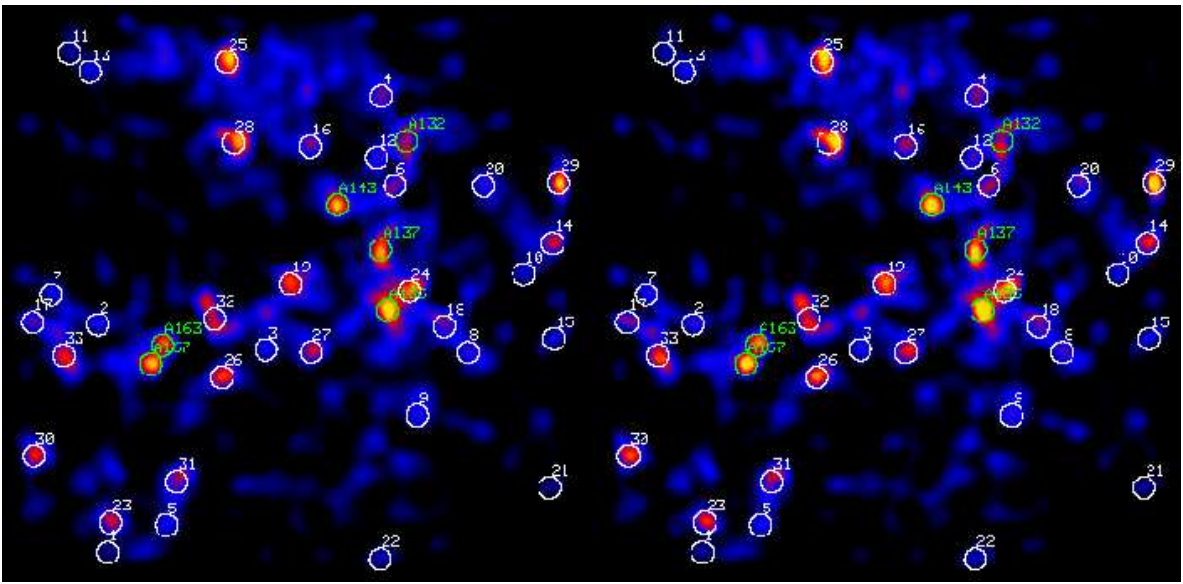}
\caption{As Figure 4, but for Field 475.}
\end{figure}

\begin{figure}
\plotone{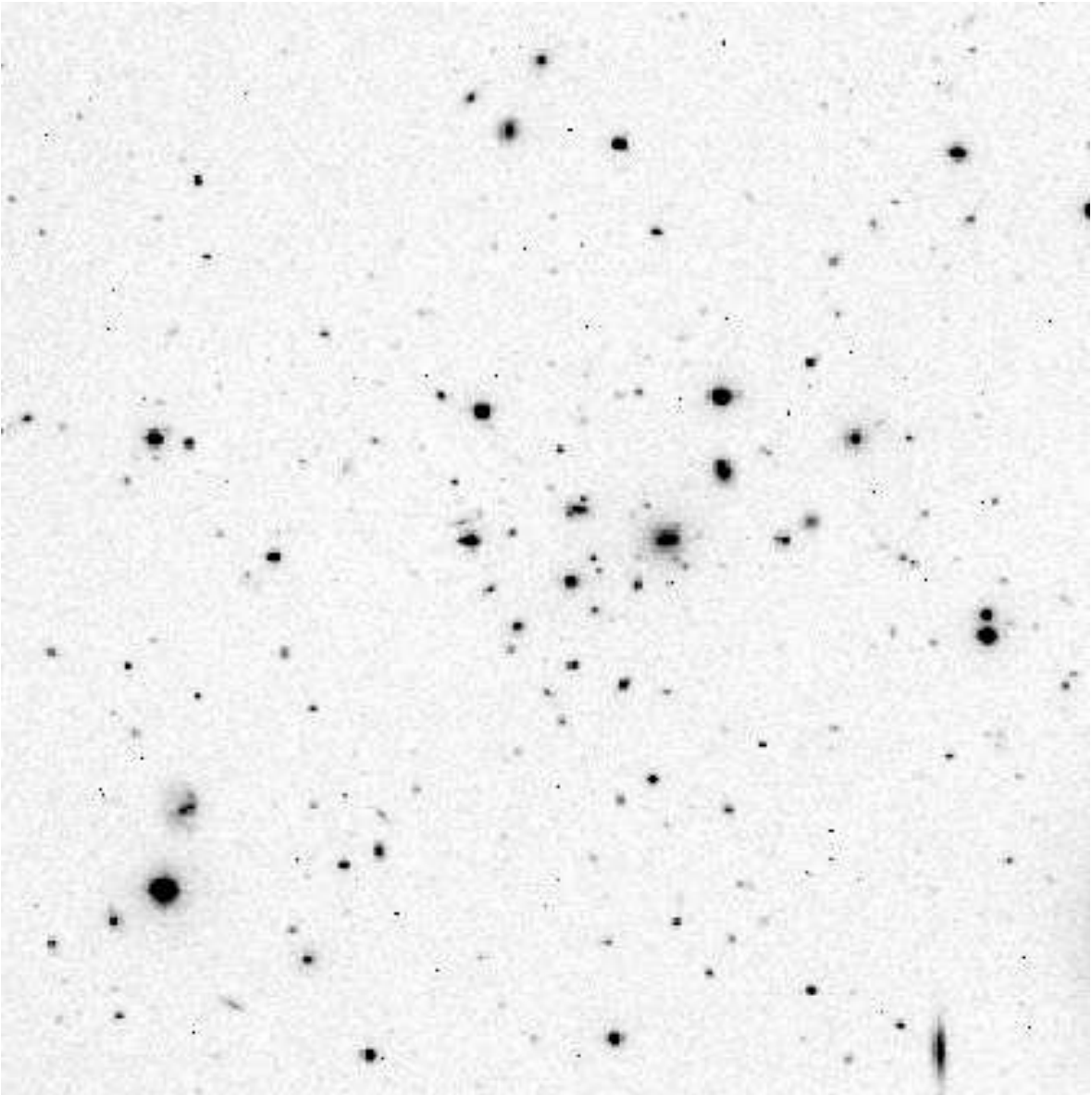}
\caption{A relatively rich cluster, candidate 31 from Field 475, missed by Abell. The photometrically estimated redshift is $z=0.193$.}
\end{figure}

\begin{figure}
\plotone{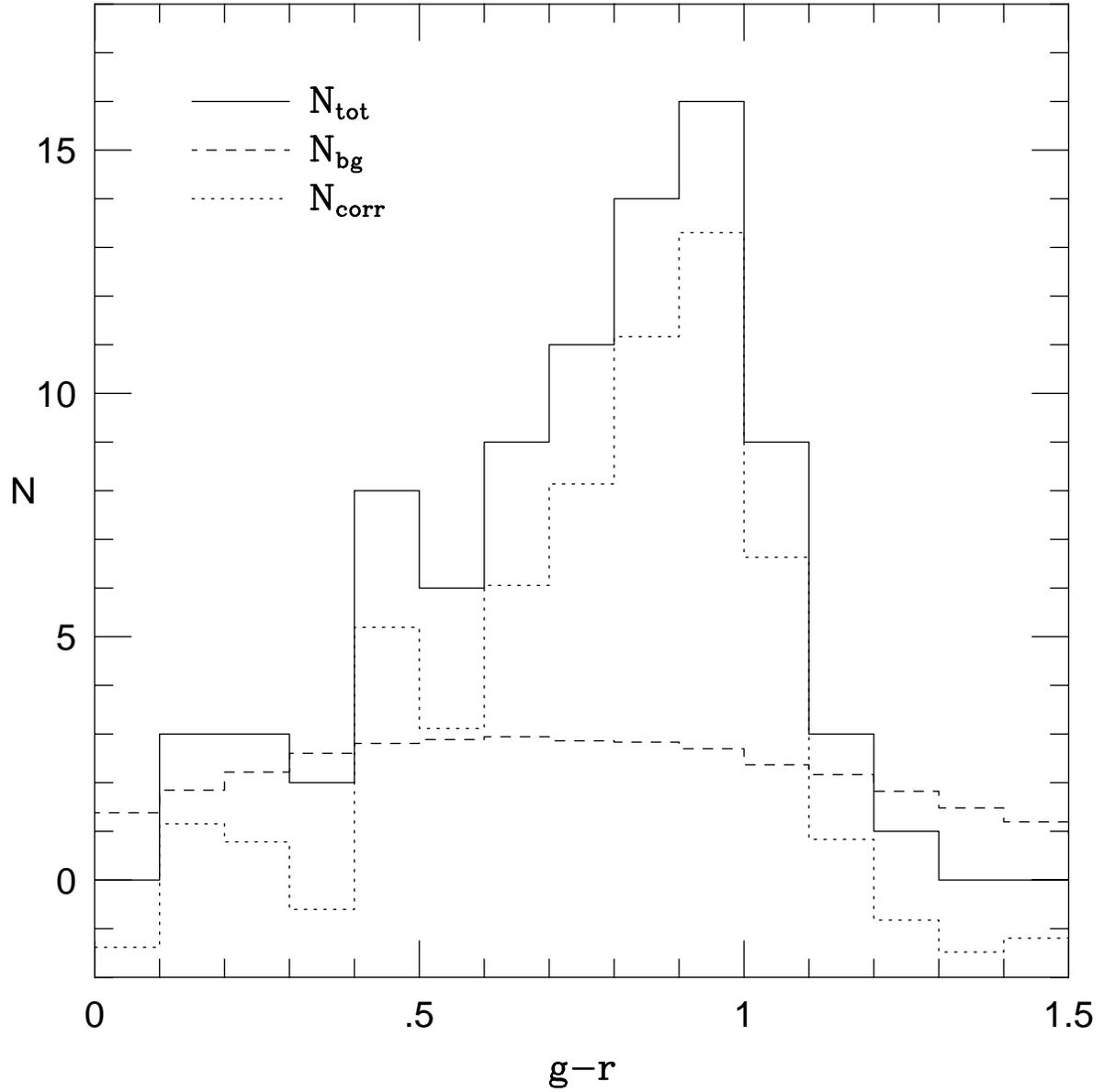}
\caption{The color distribution of Abell 115. The solid line is the 
uncorrected $N_{g-r}$. The dashed line shows the background correction, $N_{bg,g-r}$, and the dotted line is the corrected counts, $N_{corr,g-r}$.}
\end{figure}

\begin{figure}
\plotone{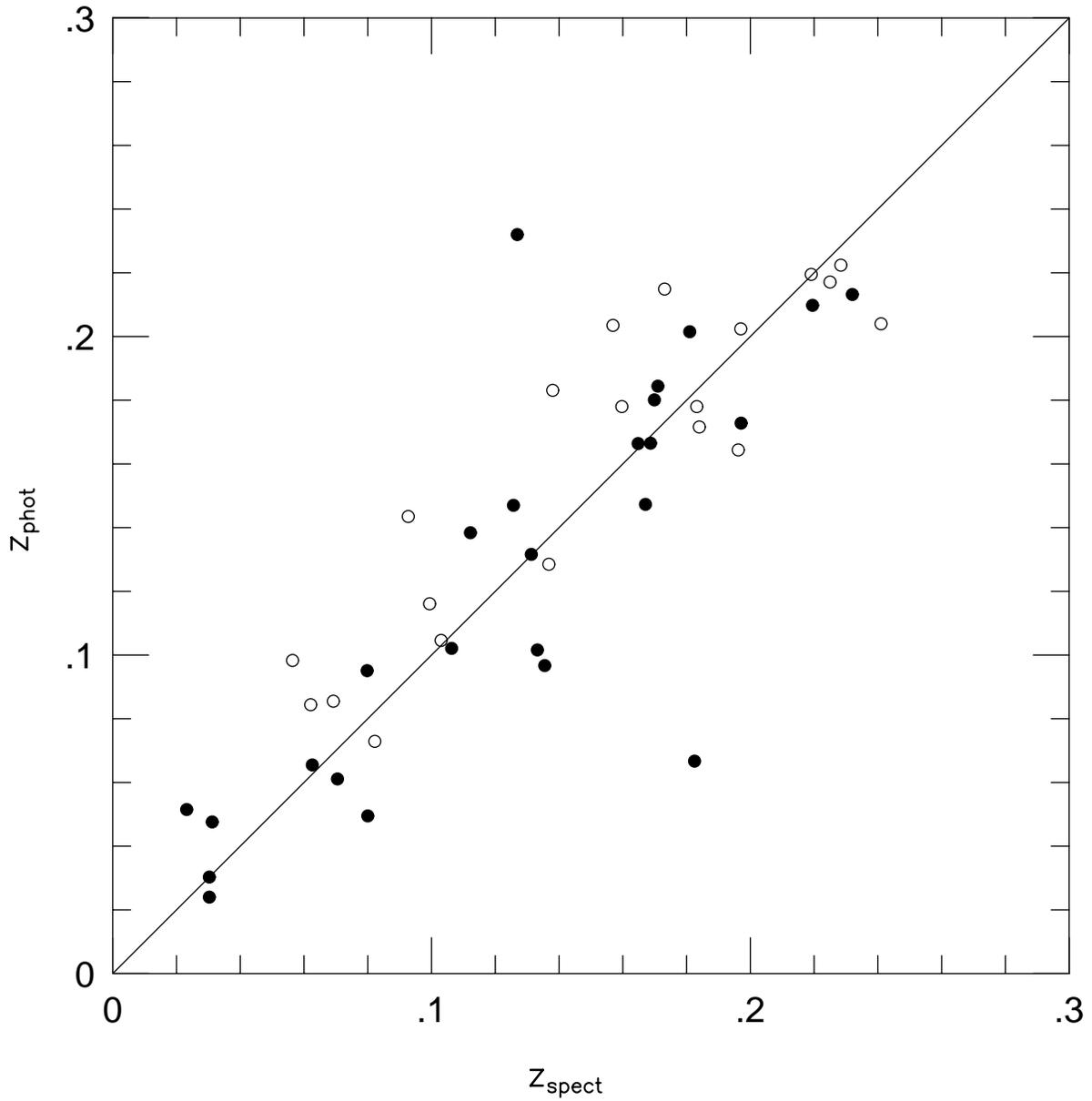}
\caption{The photometrically estimated redshift against the spectroscopically measured redshift for all 46 clusters. Open circles show those clusters with only one galaxy with a spectroscopic redshift, and filled circles are clusters with two or more spectroscopic redshifts.}
\end{figure}

\begin{figure}
\plotone{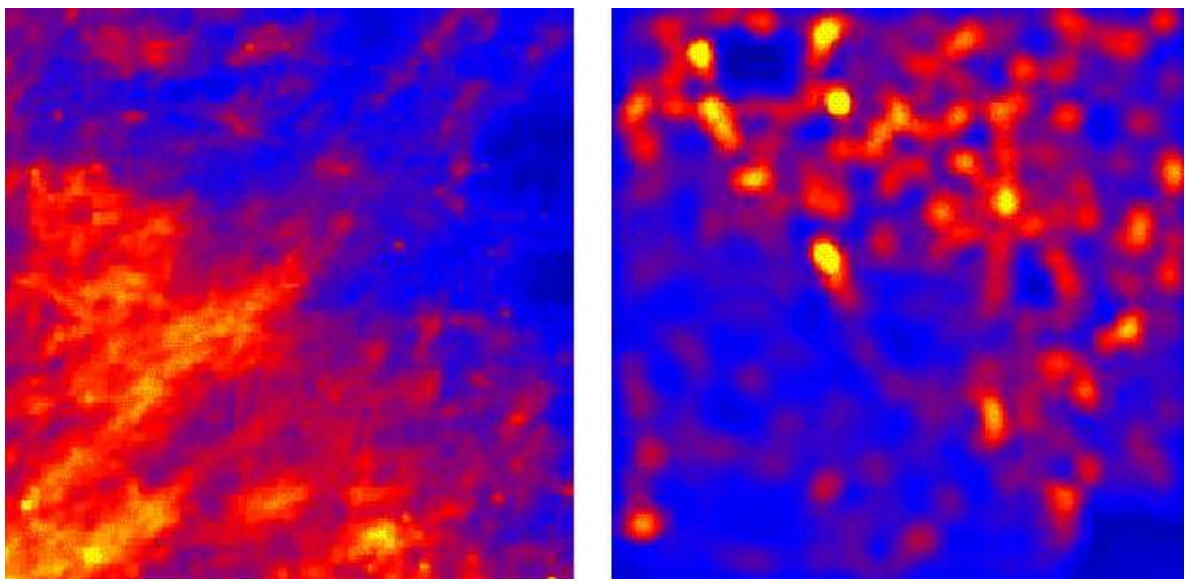}
\caption{The $100\mu$ IRAS map (left), and DPOSS galaxy density map (right), for Field 479. The strong correlation between high IR--flux and low observed galaxy density is evident.}
\end{figure}

\clearpage

\footnotesize
\begin{deluxetable}{crrrl}
\tablenum{1}
\tablecolumns{5}  
\tablewidth{0pc}  
\tablecaption{Candidate Galaxy Clusters, Field 447 \label{tbl-1}}
\tablehead{\colhead{Candidate} & \colhead{RA} & \colhead{Dec} & \colhead{$z_{gr}
$} & \colhead{Abell}}
\startdata
1 & 14 47 58.79 & +28 46 52.3 & 0.072 & -- \\
2 & 14 26 40.79 & +32 29 00.2 & 0.165 & -- \\
3 & 14 27 37.51 & +29 27 53.3 & 0.131 & -- \\
4 & 14 29 18.62 & +27 03 23.7 & 0.159 & -- \\
5 & 14 32 52.71 & +31 32 24.9 & 0.137 & 1930 \\
6 & 14 42 16.42 & +29 45 02.9 & 0.158 & -- \\
7 & 14 43 15.92 & +30 57 57.2 & 0.222 & 1958/7 \\
8 & 14 35 40.52 & +28 45 32.2 & 0.169 & -- \\
9 & 14 30 53.16 & +28 06 58.1 & 0.175 & -- \\
10 & 14 45 57.85 & +26 54 57.7 & --- & -- \\
11 & 14 41 32.63 & +26 51 57.0 & 0.139 & -- \\
12 & 14 48 28.16 & +29 58 37.4 & 0.196 & -- \\
13 & 14 46 46.24 & +29 44 22.6 & 0.120 & -- \\
14 & 14 29 41.16 & +30 13 16.8 & 0.045 & -- \\
15 & 14 23 13.26 & +32 07 38.8 & 0.128 & -- \\
16 & 14 37 34.65 & +30 08 59.3 & 0.182 & 1943 \\
17 & 14 26 40.88 & +32 11 21.3 & 0.235 & -- \\
18 & 14 42 27.50 & +32 14 15.3 & 0.185 & -- \\
19 & 14 46 19.15 & +27 17 59.4 & 0.214 & -- \\
20 & 14 46 04.66 & +31 01 30.4 & 0.240 & -- \\
21 & 14 28 50.50 & +32 30 57.7 & 0.114 & -- \\
22 & 14 39 40.49 & +30 46 10.5 & 0.125 & -- \\
23 & 14 45 07.61 & +30 00 48.2 & 0.164 & -- \\
24 & 14 44 04.05 & +30 26 23.3 & 0.234 & -- \\
25 & 14 46 55.01 & +31 51 47.8 & 0.225 & -- \\
26 & 14 39 18.80 & +27 11 57.9 & 0.077 & -- \\
27 & 14 46 44.26 & +28 21 02.3 & 0.157 & -- \\
28 & 14 42 16.84 & +29 21 41.0 & 0.207 & -- \\
29 & 14 27 47.32 & +27 42 57.4 & 0.190 & -- \\
30 & 14 34 29.26 & +28 38 40.2 & 0.162 & -- \\
31 & 14 49 12.93 & +32 33 13.9 & 0.190 & -- \\
32 & 14 43 30.10 & +31 30 49.5 & 0.142 & -- \\
33 & 14 39 17.81 & +29 01 04.7 & 0.193 & -- \\
34 & 14 30 27.37 & +30 44 40.1 & 0.167 & -- \\
35 & 14 42 34.57 & +32 31 44.8 & 0.206 & -- \\
36 & 14 43 28.24 & +27 52 27.0 & 0.237 & -- \\
37 & 14 49 03.04 & +31 48 50.9 & 0.144 & -- \\
38 & 14 48 41.06 & +30 35 45.2 & 0.138 & -- \\
39 & 14 35 45.34 & +27 57 42.4 & 0.168 & -- \\
40 & 14 32 07.91 & +29 32 57.9 & 0.229 & 1929 \\
41 & 14 45 02.37 & +31 29 49.5 & 0.194 & 1963 \\
42 & 14 37 25.40 & +30 23 50.4 & 0.119 & 1941/4 \\
43 & 14 48 30.15 & +27 24 09.5 & 0.204 & -- \\
44 & 14 42 57.74 & +31 43 50.8 & 0.173 & 1956 \\
45 & 14 44 36.71 & +31 12 16.5 & 0.215 & 1961 \\
46 & 14 33 29.08 & +29 26 25.2 & 0.207 & 1934 \\
47 & 14 41 48.43 & +28 30 25.3 & 0.155 & 1954/2 \\
\enddata
\end{deluxetable}
\clearpage
\footnotesize
\begin{deluxetable}{crrrl}
\tablenum{2}
\tablecolumns{5}  
\tablewidth{0pc}  
\tablecaption{Candidate Galaxy Clusters, Field 475 \label{tbl-2}}
\tablehead{\colhead{Candidate} & \colhead{RA} & \colhead{Dec} & \colhead{$z_{gr}
$} & \colhead{Abell}}
\startdata
1 & 01 00 30.55 & +24 37 35.8 & 0.113 & -- \\
2 & 01 13 07.97 & +22 22 07.7 & 0.174 & -- \\
3 & 01 04 33.40 & +24 36 51.8 & 0.208 & -- \\
4 & 01 21 01.38 & +22 49 16.3 & 0.163 & -- \\
5 & 01 02 03.88 & +25 00 16.3 & 0.205 & -- \\
6 & 00 59 39.76 & +26 26 20.6 & 0.187 & -- \\
7 & 01 08 01.69 & +27 25 40.8 & 0.186 & -- \\
8 & 01 19 38.48 & +27 50 28.1 & 0.228 & -- \\
9 & 01 18 30.49 & +27 36 37.5 & 0.209 & -- \\
10 & 01 03 59.18 & +26 24 41.3 & 0.200 & -- \\
11 & 00 56 38.96 & +23 11 38.7 & --- & -- \\
12 & 00 56 21.59 & +25 46 45.0 & 0.152 & -- \\
13 & 01 14 33.23 & +22 47 31.9 & 0.161 & -- \\
14 & 01 02 53.71 & +23 59 14.1 & 0.244 & -- \\
15 & 01 04 23.78 & +27 23 29.2 & 0.248 & -- \\
16 & 01 03 16.15 & +25 17 42.5 & 0.192 & -- \\
17 & 01 07 52.96 & +26 50 41.9 & 0.221 & -- \\
18 & 01 15 58.15 & +22 30 41.7 & 0.188 & -- \\
19 & 00 56 25.91 & +24 46 28.3 & 0.183 & -- \\
20 & 01 20 40.92 & +24 53 52.3 & 0.184 & -- \\
21 & 01 17 15.05 & +22 50 32.2 & 0.157 & -- \\
22 & 01 04 37.05 & +22 26 41.3 & 0.145 & -- \\
23 & 01 11 56.40 & +27 44 35.6 & 0.165 & -- \\
24 & 01 08 46.71 & +25 21 34.6 & 0.186 & -- \\
25 & 01 14 52.26 & +24 43 03.9 & 0.176 & 163 \\
26 & 01 14 06.07 & +23 15 09.0 & 0.162 & -- \\
27 & 01 04 15.75 & +25 04 51.7 & 0.200 & 136 \\
28 & 01 15 23.60 & +24 31 01.1 & 0.204 & 167 \\
29 & 01 19 25.93 & +24 32 54.5 & 0.192 & -- \\
30 & 01 06 41.84 & +26 12 58.6 & 0.208 & 143 \\
31 & 01 12 02.23 & +24 23 38.5 & 0.193 & -- \\
32 & 01 20 50.21 & +23 31 52.2 & 0.154 & -- \\
33 & 01 07 52.44 & +24 38 15.6 & 0.220 & -- \\
34 & 01 04 34.41 & +25 43 49.3 & 0.176 & 137 \\
35 & 01 11 33.55 & +26 52 55.0 & 0.169 & -- \\
36 & 00 56 01.31 & +26 25 39.0 & 0.200 & -- \\
37 & 01 03 35.36 & +26 52 21.7 & 0.142 & 132 \\
38 & 01 12 31.22 & +25 00 54.2 & 0.144 & -- \\
\enddata
\end{deluxetable}

\end{document}